\documentclass[a4paper,fleqn]{cas-sc}

\usepackage{bm}
\usepackage[numbers]{natbib}

\def\tsc#1{\csdef{#1}{\textsc{\lowercase{#1}}\xspace}}
\tsc{WGM}
\tsc{QE}

\begin{document}
\let\WriteBookmarks\relax
\def\floatpagepagefraction{1}
\def\textpagefraction{.001}
\graphicspath{ {figs/} }
\shorttitle{PINN for Heat Conduction}    

\shortauthors{Z. Tao et al.}  

\title {Analytical and Neural Network Approaches for Solving Two-Dimensional Nonlinear Transient Heat Conduction}  



%

\author[1]{Ze Tao}[orcid=0009-0004-0202-3641]
\credit{Calculation, data analyzing and manuscript writing}
\affiliation[1]{organization={Nanophotonics and Biophotonics Key Laboratory of Jilin Province, School of Physics, Changchun University of Science and Technology},
                city={Changchun},
                postcode={130022},
                country={P.R. China}}

\author[1]{Fujun Liu}[orcid=0000-0002-8573-450X]
\credit{Review and Editing}
\cormark[1]
\ead{fjliu@cust.edu.cn}
\cortext[1]{Corresponding author}

\author[1]{Jinhua Li}[orcid=0000-0003-3505-6808]
\credit{Review and Editing}

\author[1]{Guibo Chen}[orcid=0000-0001-8436-2284]
\credit{Review and Editing}

\begin{abstract}
Accurately predicting nonlinear transient thermal fields in two-dimensional domains is a significant challenge in various engineering fields, where conventional analytical and numerical methods struggle to balance physical fidelity with computational efficiency when dealing with strong material nonlinearities and evolving multiphysics boundary conditions. To address this challenge, we propose a novel cross-disciplinary approach integrating Green’s function formulations with adaptive neural operators, enabling a new paradigm for multiphysics thermal analysis. Our methodology combines rigorous analytical derivations with a physics-informed neural architecture consisting of five adaptive hidden layers (64 neurons per layer) that incorporates solutions as physical constraints, optimizing learning rates to balance convergence stability and computational speed. Extensive validation demonstrates superior performance in handling rapid thermal transients and strongly coupled nonlinear responses, which significantly improves computational efficiency while maintaining high agreement with analytical benchmarks across a range of material configurations and boundary conditions. 
\end{abstract}


\begin{highlights}
\item Analytical approaches integrated with adaptive neural operators enable the prediction of nonlinear thermal field.
\item The physics-informed network, consisting of 5×64 adaptive layers, balances convergence stability and computational speed.
\item It  maintains high agreement with analytical benchmarks across material configurations and thermal regimes.
\item The network is able to resolve strong material nonlinearities and evolving multiphysics boundary conditions.
\item Validation demonstrates efficient handling of rapid transients and coupled nonlinear responses.
\end{highlights}


\maketitle

\section{Introduction}
The analysis of two-dimensional nonlinear transient heat conduction problems is crucial in modern engineering applications \cite{ref9,ref11,ref30}, such as aerospace thermal protection systems \cite{ref7} and electronic device thermal management \cite{ref8}. As the demand for high-precision temperature field predictions under extreme operating conditions increases, accurate modelling of these phenomena has become essential for reliability assessment and structural optimization. However, existing solution methodologies often struggle to balance computational efficiency\cite{ref31} with physical accuracy, particularly when dealing with nonlinear material properties, time-dependent boundary conditions, and complex geometric configurations. Traditional numerical approaches \cite{ref13,ref14, ref15}, such as finite element methods and finite volume schemes, frequently encounter challenges like excessive computational burden in long-term transient analyses and convergence instability in strongly nonlinear regimes, especially when addressing multi-scale heat transfer phenomena \cite{ref16}. This gap between theoretical modelling needs and practical computational capabilities underscores the urgent necessity for hybrid methodologies that combine rigorous mathematical analysis with modern computational intelligence techniques.

Recent advancements in computational heat transfer have made notable strides in solving transient heat conduction problems through both analytical and numerical frameworks \cite{ref17,ref29}. Classical methods, such as Green’s function-based techniques and separation of variables, have been widely employed to derive closed-form solutions for simplified geometries and linear boundary conditions \cite{ref19}. On the other hand, finite element methods (FEM) \cite{ref20,ref32} and finite volume schemes \cite{ref21} have proven effective in solving nonlinear problems, offering flexibility in handling irregular domains and temperature-dependent material properties. Despite these developments, critical limitations persist: analytical methods often rely on simplifying assumptions (e.g., linearized source terms or idealized boundary conditions) to maintain tractability, while purely numerical methods face prohibitive computational costs when resolving fine temporal-spatial scales in long-duration transient analyses \cite{ref22}. Hybrid strategies, such as combining perturbation techniques with spectral methods, have shown some potential to alleviate these issues but remain insufficient for handling strongly coupled nonlinear systems \cite{ref23}.

In response to these limitations, emerging machine learning applications, particularly physics-informed neural networks (PINNs) \cite{ref24}, have garnered attention for their potential to accelerate heat transfer simulations. However, many existing implementations overlook systematic investigations into hyperparameter sensitivity \cite{ref25}, such as optimizing learning rates, and fail to rigorously validate results against high-fidelity analytical benchmarks. This compromises their reliability, especially in mission-critical engineering contexts. The integration of analytical rigor with data-driven computational techniques, through hybrid methodologies, presents a transformative opportunity to address the shortcomings of traditional nonlinear heat conduction analysis. Recent advances in physics-informed machine learning frameworks have successfully embedded fundamental conservation laws directly into neural network architectures \cite{ref26}, ensuring physical consistency while benefiting from the flexibility of deep learning.

A particularly promising approach is the combination of Green’s function solutions with adaptive neural operators. This integration can preserve mathematical exactness in linear regimes while efficiently approximating nonlinear corrections. By exploiting closed-form solutions for baseline predictions and deploying neural networks for residual modelling, this approach balances computational efficiency with physical fidelity. Moreover, optimizing critical hyperparameters such as learning rate scheduling and network depth has proven effective in stabilizing the training process and improving solution accuracy \cite{ref27}. This paradigm shift bridges the gap between traditional analytical methods and modern computational intelligence, offering a unified framework capable of handling complex boundary conditions, material nonlinearities, and multi-physics coupling effects that have long challenged conventional approaches.

In this work, we introduce a novel dual-methodology framework that synergizes analytical rigor with adaptive neural network modelling, thereby addressing both theoretical and computational gaps in nonlinear transient heat conduction analysis. Our approach systematically integrates pulse decomposition (Green’s function) and integral transform methods, establishing a unified analytical foundation and rigorously proving their mathematical equivalence. We demonstrate that integral transforms outperform Green’s function-based methods in practical applications, particularly when dealing with spatially varying source terms and non-homogeneous boundary conditions. A physics-informed neural network architecture with optimized hyperparameters is developed by addressing the critical relationship between learning rate dynamics and model performance. Through exhaustive parametric studies, we identify an optimal learning rate configuration (0.005) that achieves an unprecedented balance between convergence stability and computational efficiency, resolving the issues of sluggish convergence or oscillatory divergence typically observed in conventional machine learning implementations \cite{ref28}. The proposed hybrid methodology not only validates its accuracy against high-fidelity analytical benchmarks but also shows remarkable adaptability to complex engineering scenarios. It intelligently decomposes the solution domain by applying analytical methods to linear subregions and neural networks to nonlinear residuals. This approach establishes a new paradigm for multi-physics simulations, reconciling the precision of mathematical analysis with the computational agility of deep learning. As a result, it significantly advances the state-of-the-art in predictive modelling of transient thermal phenomena, with implications for a wide range of engineering applications.

\section{Problem setup}
Consider a two-dimensional rectangular region, denoted as $\left\{\left(x,y,t\right)\in R^{2}\times [0,+\infty)\,|\,0\leq x\leq L_{x},\,0\leq y\leq L_{y},\,t\geq 0\right\}$,  which is situated within the plane $O_{xy}$. It is essential to define the boundary conditions applied to this region. Specifically, the boundary temperature is set to zero and the initial temperature throughout the entire region is also zero. Therefore, the governing equation for the internal heat source within this region can be expressed as:
\begin{subequations}\label{2.1.1}
	\begin{align}
		&\frac{\partial^{2}T}{\partial x^{2}}+\frac{\partial^{2}T}{\partial y^{2}}+\frac{g\left(x,y,t\right)}{k}=\frac{1}{\alpha}\frac{\partial T}{\partial t}\quad t\geq0, 0\leq x\leq L_{x}, 0\leq y\leq L_{y}\\
		B.C.\quad&T=0,\quad t>0,\quad \text{Total boundary}\\
		I.C.\quad&T=0,\quad t=0,\quad \text{Total region}
	\end{align}
\end{subequations}
where the term $k$ denotes the thermal conductivity, the symbol $\alpha$ represents the thermal diffusion coefficient, and the symbol $g\left(x,y,t\right)$ denotes the internal heat source term. 

\section{General Solution of the two-dimensional Nonlinear Transient Heat Conduction Problem}
The results obtained using the Green's function method and the integral transformation method are presented below. For more information on the derivation details, please refer to the appendix.
\subsection{The solution by Green's function method}
The general three-dimensional nonlinear transient heat conduction problem can be formulated as follows:
\begin{subequations}\label{eqn-1a}
	\begin{align}
		&\nabla^{2}T\left(\bf{r},\mathnormal{t}\right)+\frac{g\left(\bf{r},\mathnormal{t}\right)}{k}=\frac{1}{\alpha}\frac{\partial T\left(\bf{r},\mathnormal{t}\right)}{\partial t},\in R (t\geq0,\in R)\label{3.1.1a}\\
		B.C.\quad &k_{i}\frac{\partial T}{\partial n_{i}}+h_{i}T=f_{i}\left(\bf{r},\mathnormal{t}\right),\quad\in S_{i} (i=1,2,...,s)\label{3.1.1b}\\
		I.C.\quad &T=F\left(\bf{r}\right),\quad\in R \label{3.1.1c}
	\end{align}
\end{subequations}
and the general solution is:
\begin{equation}
	T\left(\bm{r},t\right)=\sum_{m=1}^{\infty}\frac{e^{-\alpha\lambda_{m}^{2}t}}{N\left(\lambda_{m}\right)}\psi\left(\lambda_{m},\bm{r}\right)\left(I_{a}+I_{b}\right),\label{3.1.56}
\end{equation}
with the integrals:
\begin{subequations}\label{3.1.57}
	\begin{align}	&I_{a}=\int_{R}\psi\left(\lambda_{m},\bm{r}^{\prime}\right)F\left(\bm{r}^{\prime}\right)\mathrm{d}V^{\prime},\\
		&I_{b}=\int_{0}^{t}e^{\alpha\lambda_{m}^{2}t^{\prime}}\left(\frac{\alpha}{k}I_{b1}+\alpha I_{b2}\right)\mathrm{d}t^{\prime},\\
		&I_{b1}=\int_{R}\psi\left(\lambda_{m},\bm{r}^{\prime}\right)g\left(\bm{r}^{\prime},t^{\prime}\right)\mathrm{d}V^{\prime},\\
		&I_{b2}=\sum_{i=1}^{s}\int_{S_{i}}\frac{\psi\left(\lambda_{m},\bm{r}^{\prime}\right)}{k_{i}}f_{i}\left(\bm{r}^{\prime},t^{\prime}\right)\mathrm{d}S_{i}.
	\end{align}
\end{subequations}

To solve the subsequent problem of Eq. \eqref{2.1.1},  we make the following substitutions to Eq. \eqref{3.1.56}:
\begin{subequations}
	\begin{align}
		&\psi\left(\lambda_{m},\bm{r}\right)\to X\left(\beta_{m},x\right)Y\left(\gamma_{n},y\right),\\
		&N\left(\lambda_{m}\right)\to N\left(\beta_{m}\right)N\left(\gamma_{n}\right),\\
		&\lambda_{m}^{2}\to\beta_{m}^{2}+\gamma_{n}^{2},\\	&\sum_{m=1}^{\infty}\to\sum_{m=1}^{\infty}\sum_{n=1}^{\infty},\\		&\int_{R}\mathrm{d}V\to\int_{0}^{L_{x}}\mathrm{d}x^{\prime}\int_{0}^{L_{y}}\mathrm{d}y^{\prime},
		\end{align}
\end{subequations}
then we have the general  solution of Eq. \eqref{2.1.1}:
\begin{equation}
    T(x,y,t) = \sum_{m=1}^{\infty}\sum_{n=1}^{\infty}\frac{\alpha }{k}\frac{X(\beta_{m},x)Y(\gamma_{n},y)}{N(\beta_{m})N(\gamma_{n})}e^{-\alpha(\beta_{m}^{2}+\gamma_{n}^{2})t}\int_{0}^{t}e^{\alpha(\beta_{m}^{2}+\gamma_{n}^{2})t^{\prime}}\overline{\overline{g}}(\beta_{m},\gamma_{n},t^{\prime})\mathrm{d}t^{\prime},\label{3.1.60}
\end{equation}
where the double transformation $\overline{\overline{g}}$ is expressed as:
\begin{equation}\label{9}
\overline{\overline{g}}(\beta_{m},\gamma_{n},t) = \int_{0}^{L_{x}}\int_{0}^{L_{y}}X(\beta_{m},x^{\prime})Y(\gamma_{n},y^{\prime}) g(x^{\prime},y^{\prime},t^{\prime})\mathrm{d}x^{\prime}\mathrm{d}y^{\prime},
\end{equation}

Therefore, the corresponding eigenfunctions, modes, and eigenvalues of Eq. \eqref{3.1.60} are given by:
\begin{subequations}
    \begin{align}
	&\small	X(\beta_{m},x)=\sin\beta_{m}x,\quad N(\beta_{m}) = \frac{L_{x}}{2},\quad \beta_{m} = \frac{m\pi}{L_{x}};\\
	&\small	Y(\gamma_{n},y)=\sin\gamma_{n}y,\quad N(\gamma_{n}) = \frac{L_{y}}{2},\quad \gamma_{n} = \frac{n\pi}{L_{y}}
    \end{align}
\end{subequations}

\subsection{The solution by integral transformation method}
For the general three-dimensional nonlinear transient heat conduction problem in Eq. \eqref{eqn-1a}, we employ the integral transformation method and  obtain the solution:
\begin{equation} 
T\left(\bm{r},t\right)=\sum_{m=1}^{\infty}\frac{\psi\left(\lambda_{m},\bm{r}\right)}{N\left(\lambda_{m}\right)}\left(\int_{0}^{t}e^{\alpha\lambda_{m}^{2}t^{\prime}}\alpha\left(\frac{1}{k}\overline{g}\left(\lambda_{m},t^{\prime}\right)+\sum_{i=1}^{s}\int_{S_{i}}\frac{\psi\left(\lambda_{m},\bm{r}\right)}{k_{i}}f_{i}\left(\bm{r},t^{\prime}\right)\mathrm{d}S_{i}\right)\mathrm{d}t^{\prime}+\overline{F}\left(\lambda_{m}\right)\right)e^{-\alpha\lambda_{m}^{2}t},\label{3.2.74}
\end{equation}
with the items:
\begin{subequations}
    \begin{align}
&\overline{F}\left(\lambda_{m}\right)=\int_{R}\psi\left(\lambda_{m},\bm{r}^{\prime}\right)F\left(\bm{r}^{\prime}\right)\mathrm{d}V^{\prime},\\    &N\left(\lambda_{m}\right)=\int_{R}\left(\psi\left(\lambda_{m},\bm{r}^{\prime}\right)\right)^{2}\mathrm{d}V^{\prime},\\ &\overline{g}\left(\lambda_{m},t^{\prime}\right)=\int_{R}\psi\left(\lambda_{m},\bm{r}^{\prime}\right)g\left(\bm{r}^{\prime},t^{\prime}\right)\mathrm{d}V^{\prime}
    \end{align}
\end{subequations}

This result aligns with the numerical solution derived from Eq. \eqref{3.1.56}.



\section{PINN solution to the two-dimensional Nonlinear Transient Heat Conduction Problem}
\subsection{Setting up of loss function }
For two-dimensional heat conduction, the loss function can be written as:
\begin{equation}
	L\left(\theta\right)=\lambda_{1}L_{1}\left(\theta\right)+\lambda_{2}L_{2}\left(\theta\right)+\lambda_{3}L_{3}\left(\theta\right),
\end{equation}
with the items
\begin{subequations}
    \begin{align}
    &L_{1}\left(\theta\right)=\frac{1}{N}\sum_{i=1}^{N}\left(\nabla^{2}T\left(\bf{r},\mathnormal{t}\right)+\frac{g\left(\bf{r},\mathnormal{t}\right)}{k}-\frac{1}{\alpha}\frac{\partial T\left(\bf{r},\mathnormal{t}\right)}{\partial t}\right)^{2},\\
    &L_{2}\left(\theta\right)=\frac{1}{M} \sum_{i=1}^M \sum_{j=1}^{M}\left[k_i \frac{\partial T}{\partial n_i}\left(\mathbf{r}_{i j}, t_{i j}\right)+h_i T\left(\mathbf{r}_{i j}, t_{i j}\right)-f_i\left(\mathbf{r}_{i j}, t_{i j}\right)\right]^2,\\
    &L_{3}\left(\theta\right)=\frac{1}{K} \sum_{i=1}^K\left[T\left(\mathbf{r}_i, 0\right)-F\left(\mathbf{r}_i\right)\right]^2 ,
    \end{align}
\end{subequations}
where $\lambda_{1/2/3}$ is the weight coefficient;  $N$, $M$ and $K$ are the numbers of samples inside the selected region, the boundary conditions and the initial moment, respectively; $L_1$, $L_2$ and $L_3$ are the residuals of the control equations, boundary conditions and the initial moment,  respectively.\\
\subsection{PINN structure and iterative process}
\begin{figure}[htbp]
	\centering
	\includegraphics[scale=0.35]{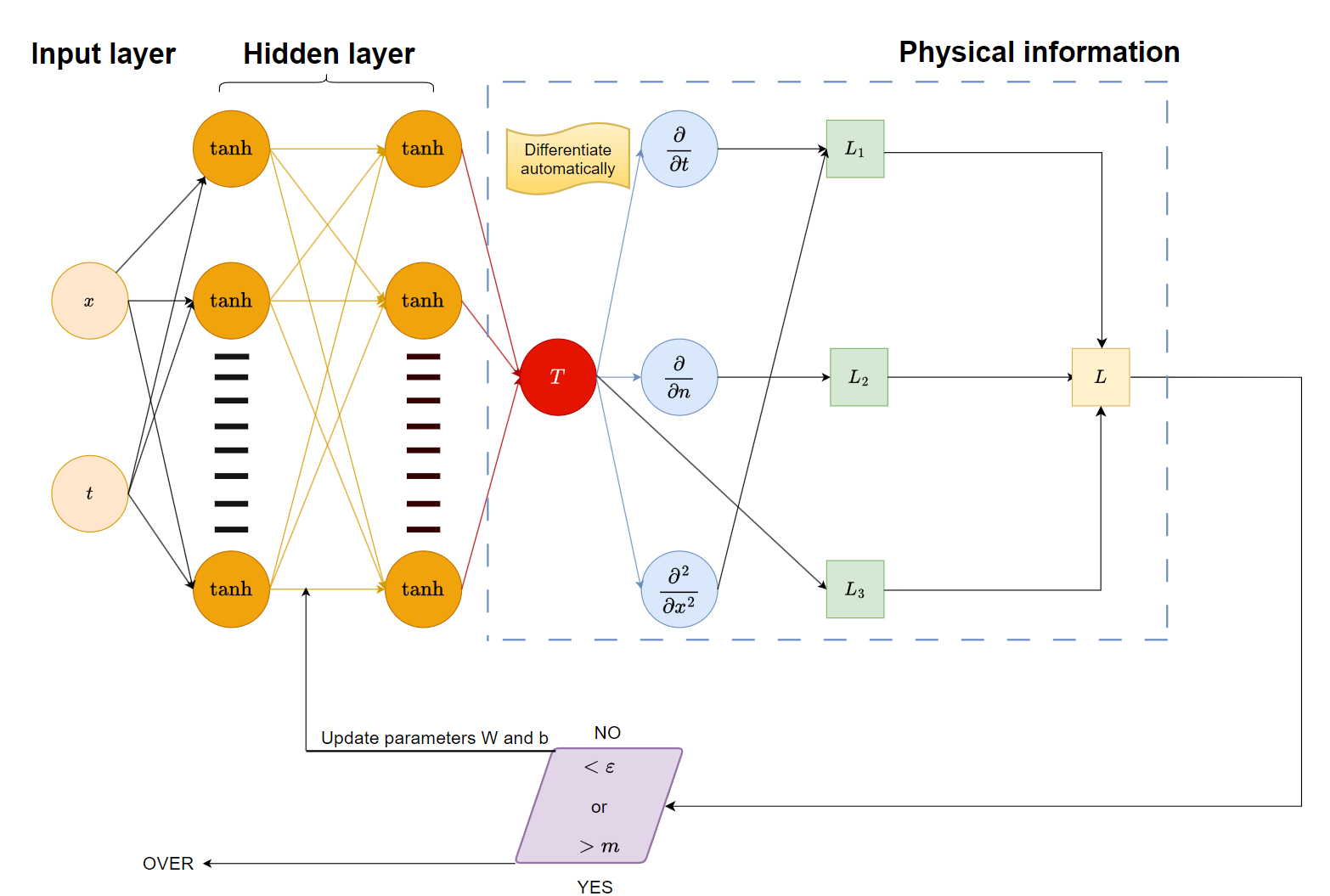}
	\caption{PINN structure diagram}
	\label{pinn}
\end{figure}

As demonstrated in the accompanying Fig. \ref{pinn}, the function is initially approximated by a fully connected neural network. With subsequent input of the temporal and spatial data, the residuals of the partial differential equations are derived via automatic differentiation techniques. The initial and marginal residual constraints are incorporated into the loss function as regular terms. Ultimately, the loss function is continuously optimized to yield the final prediction value. The stochastic gradient descent (SGD) method is utilized to continuously update and obtain the optimal parameters $\theta=\left[W,b\right]$ of the neural network, and the temperature distribution of the entire field is predicted by the PINN model. 

It has been established that, within the context of the PINN training process, the loss function can be configured to terminate training when the loss function is less than a specified value of $\varepsilon$, or when the number of iterations is less than a specified value of $m$.

\section{Numerical examples and analytical solutions to specific problems}
\subsection{Series solution to the two-dimensional nonlinear transient heat conduction problem}
Taking a hard material tungsten carbide (WC) as an example, with a thermal conductivity $k$ of 75W/(m$\cdot$K), thermal diffusivity $\alpha$ of $2.5\times10^{-5} \, \mathrm{m^{2} / s}$, the lengths of $L_{x}$ = 0.4 mm  and $L_{y}$ = 2.5 mm. The general solution for the two-dimensional nonlinear heat conduction problem \cite{ref1} is:
\begin{equation}\label{4}
	T(x,y,t) = \sum_{m=1}^{\infty}\sum_{n=1}^{\infty}\frac{\alpha }{k}\frac{X(\beta_{m},x)Y(\gamma_{n},y)}{N(\beta_{m})N(\gamma_{n})}e^{-\alpha(\beta_{m}^{2}+\gamma_{n}^{2})t}\int_{0}^{t}e^{\alpha(\beta_{m}^{2}+\gamma_{n}^{2})t^{\prime}}\overline{\overline{g}}(\beta_{m},\gamma_{n},t^{\prime})\mathrm{d}t^{\prime},
\end{equation}
where
\begin{equation}\label{7}
	g(x,y,t) = \dot{q}(x,y,t),
\end{equation}
and
\begin{equation}\label{8}
\dot{q}(x,y,t) = q^{\prime\prime}\left(t\right)\left[C_{1} + \frac{36(1 - C_{1})}{L_{x}^{2}L_{y}^{2}}xy(L_{x} - x)(L_{y} - y)\right]
\end{equation}

The double transformation is given by:
\begin{equation}\label{9}
	\overline{\overline{g}}(\beta_{m},\gamma_{n},t) = \int_{0}^{L_{x}}\int_{0}^{L_{y}}X(\beta_{m},x^{\prime})Y(\gamma_{n},y^{\prime}) g(x^{\prime},y^{\prime},t^{\prime})\mathrm{d}x^{\prime}\mathrm{d}y^{\prime},
\end{equation}
which has the eigenfunctions, modes, and eigenvalues as follow:
\begin{equation}\label{10}
	X(\beta_{m},x)=\sin\beta_{m}x,\quad N(\beta_{m}) = \frac{L_{x}}{2},\quad \beta_{m} = \frac{m\pi}{L_{x}};
\end{equation}
\begin{equation}\label{11}
Y(\gamma_{n},y)=\sin\gamma_{n}y,\quad N(\gamma_{n}) = \frac{L_{y}}{2},\quad \gamma_{n} = \frac{n\pi}{L_{y}}
\end{equation}

Substituting $C_{1}$ = 0.5, $q^{\prime\prime}$(t) = 30MW/$m^2$, conductivity $k$ and thermal diffusivity $\alpha$ into equations \eqref{8} and \eqref{9}, the double transformation becomes:
\begin{equation}
	\overline{\overline{g}}(\beta_{m},\gamma_{n},t) = \int_{0}^{L_{x}}\int_{0}^{L_{y}}\sin\left(\beta_{m}x^{\prime}\right)\sin\left(\gamma_{n}y^{\prime}\right)(15 + 540x^{\prime}y^{\prime}(L_{x} - x^{\prime})(L_{y} - y^{\prime}))\mathrm{d}x^{\prime}\mathrm{d}y^{\prime},
\end{equation}

Thus, $\overline{\overline{g}}(\beta_{m},\gamma_{n},t)$
is independent of time, and after computation, the result is:
\begin{equation}\label{16}
	\overline{\overline{g}}(\beta_{m},\gamma_{n}) = \frac{15}{\beta_{m}\gamma_{n}}(1 - \cos\beta_{m}L_{x})(1 - \cos\gamma_{n}L_{y}) + 540(-\frac{L_{x}}{\beta_{m}^{2}}\sin\beta_{m}L_{x} + \frac{2}{\beta_{m}^{3}}(1 - \cos\beta_{m}L_{x}))(-\frac{L_{y}}{\gamma_{n}^{2}}\sin\gamma_{n}L_{y} + \frac{2}{\gamma_{n}^{3}}(1 - \cos\gamma_{n}L_{y})),
\end{equation}

Substituting equation \eqref{16} into equation \eqref{4}, the result is:
\begin{equation}\label{21}
    T(x,y,t) = \sum_{m=1}^{\infty}\sum_{n=1}^{\infty}\frac{\alpha }{k}\frac{X(\beta_{m},x)Y(\gamma_{n},y)}{N(\beta_{m})N(\gamma_{n})}\frac{\overline{\overline{g}}(\beta_{m},\gamma_{n})}{\alpha(\beta_{m}^{2} + \gamma_{n}^{2})}\left(1 - e^{-\alpha(\beta_{m}^{2}+\gamma_{n}^{2})t}\right),
\end{equation}

Substituting equations \eqref{10} and \eqref{11}  into equation \eqref{21}, we obtain:
\begin{equation}\label{22}
	T(x,y,t) = \sum_{m=1}^{\infty}\sum_{n=1}^{\infty}\frac{4\sin(\beta_{m}x)\sin(\gamma_{n}y)\overline{\overline{g}}(\beta_{m},\gamma_{n})}{\alpha(\beta_{m}^{2}+\gamma_{n}^{2})L_{x}L_{y}}\left(1 - e^{-\alpha(\beta_{m}^{2}+\gamma_{n}^{2})t}\right),
\end{equation}
where $\overline{\overline{g}}(\beta_{m},\gamma_{n})$ maintains equation \eqref{16}.

\begin{figure}[htbp]
	\centering
	\includegraphics[scale=0.25]{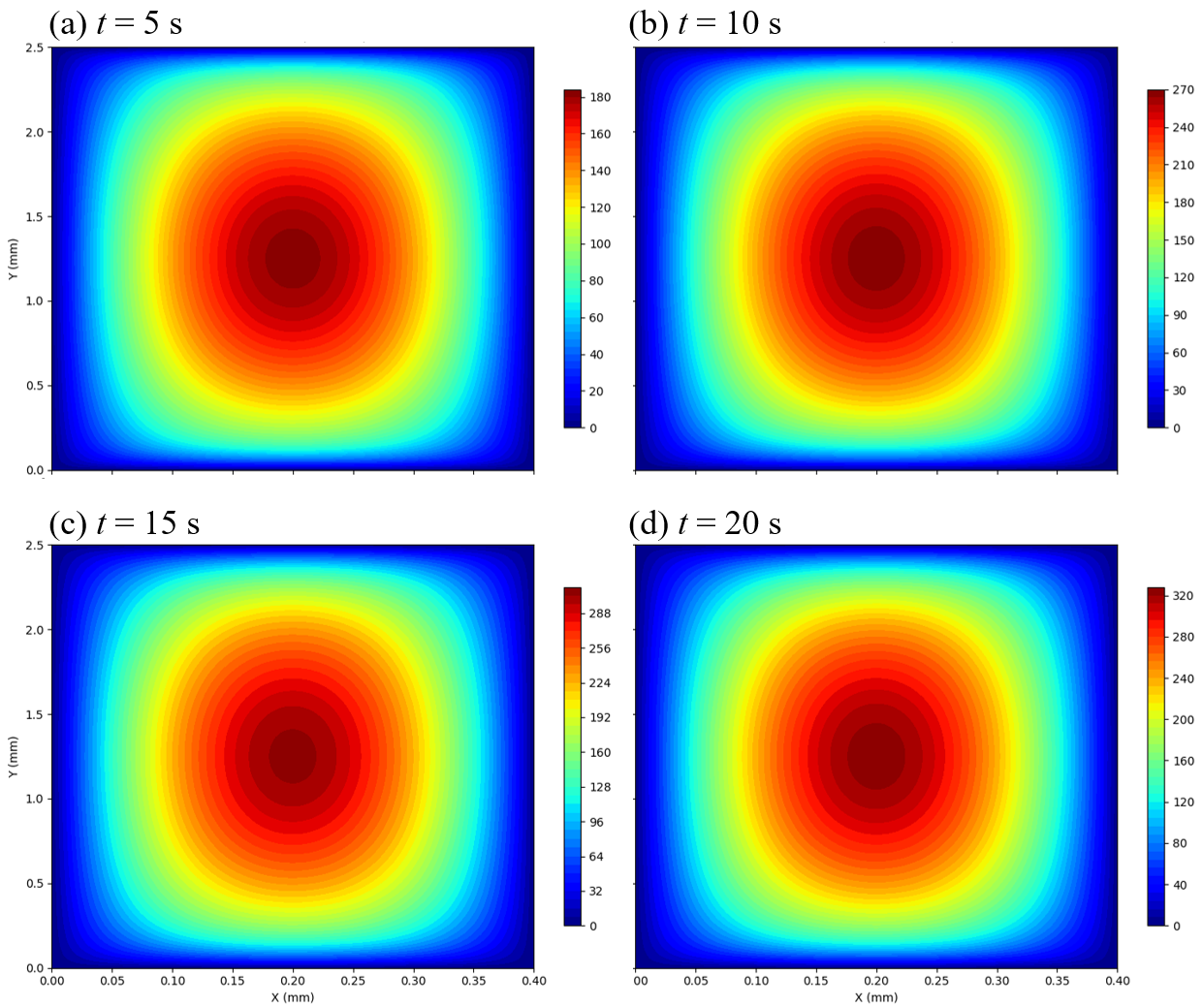}
	\caption{Results for the analytical solution at various times ($t$).}.
	\label{analytical solution}
\end{figure}

Equation \eqref{22} represents the analytical solution to the two-dimensional nonlinear transient heat conduction problem.And the temperature distribution of this solution is shown in Fig. \ref{analytical solution}.

\subsection{The influence of various learning rates on computational results}
This study investigates the effect of varying learning rates on the performance of a neural network model. The model consists of five hidden layers, each containing 64 neurons. The input layer receives three features $\left(x,y,t\right)$, and the output layer has a single neuron that predicts the temperature. The Adam optimizer is employed, which adaptively adjusts the learning rate to enhance the training efficiency \cite{refAdam}. To introduce non-linearity and improve the model’s representational capacity, the Tanh activation function is applied to each hidden layer \cite{Tanh}. PyTorch automatically initializes the weights \cite{pytorch}, and backpropagation is used to optimize the model, gradually updating the parameters to improve performance. 

\begin{figure}[htbp]
	\centering
	\includegraphics[scale=0.25]{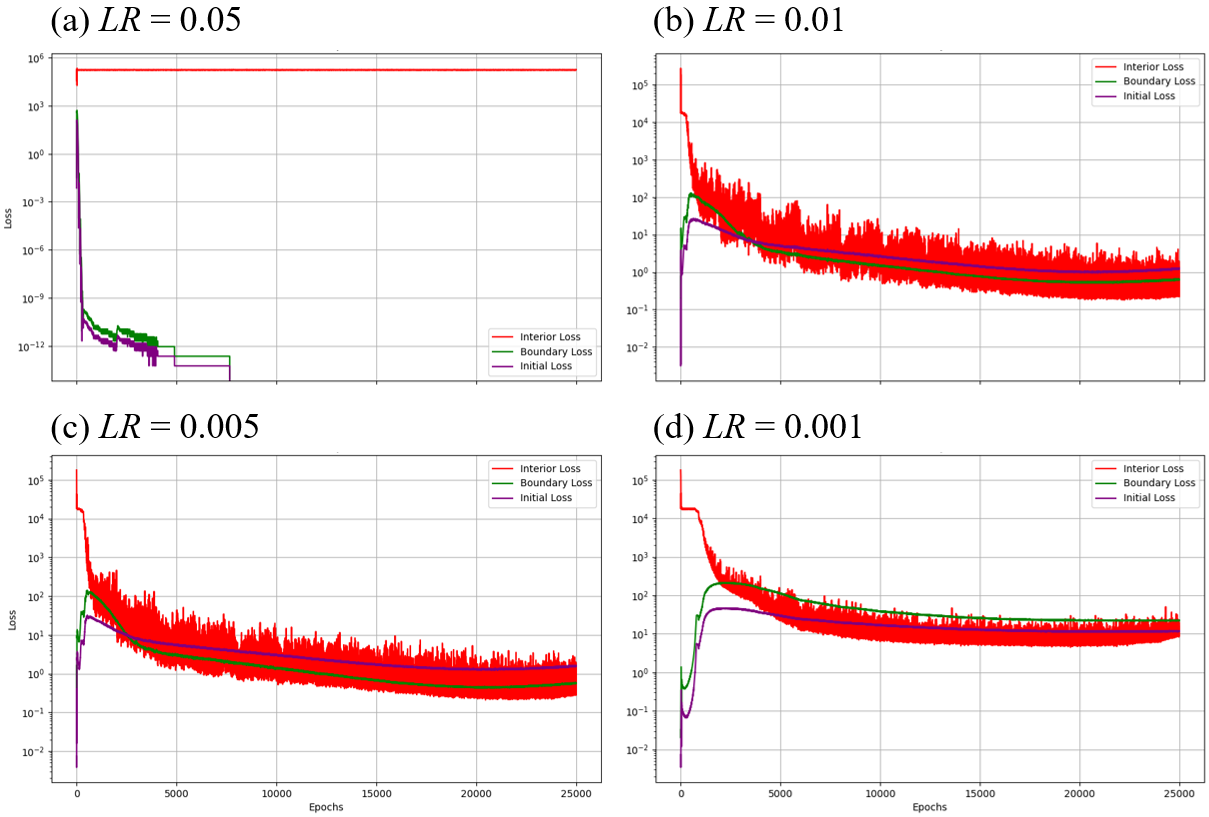}
	\caption{The loss function with different learning rate ($LR$).}
	\label{F1}
\end{figure}

Figure \ref{F1} illustrate how different learning rates affect the model’s convergence speed and stability. With a learning rate of 0.001, the model eventually converges but the training process is slow due to the small step size, and the loss function decreases minimally. This sluggish pace may cause the model to become trapped in a local minimum, limiting its ability to effectively explore the solution space. Early training struggles with oscillations, leading to inefficient convergence. Increasing the learning rate to 0.005 improves both convergence speed and stability. Compared to the 0.001 rate, the loss function decreases more rapidly with smaller oscillations, enabling the model to converge more quickly to a better solution. This learning rate strikes an optimal balance between training efficiency and stability, making it the most effective choice. With a learning rate of 0.01, the loss function decreases even more quickly, but oscillations become more pronounced. While stability is slightly worse than with a learning rate of 0.005 in some instances, the model still maintains acceptable stability, making this learning rate suitable for situations where faster convergence is prioritized. With a learning rate of 0.05, the loss function decreases rapidly but with significant oscillations, leading to instability during training. The large step size hampers stable convergence and may even cause gradient explosion, ultimately degrading model performance. In conclusion, a learning rate that is too low (e.g., 0.001) leads to slow convergence and risks trapping the model in a local optimum. On the other hand, a learning rate that is too high (e.g., 0.05) causes instability, negatively affecting the final outcomes. Learning rates of 0.005 and 0.01 offer a favorable trade-off between efficiency and stability, with 0.005 being the most effective in balancing training speed and result accuracy. 

\begin{figure}[htbp]
	\centering
	\includegraphics[scale=0.38]{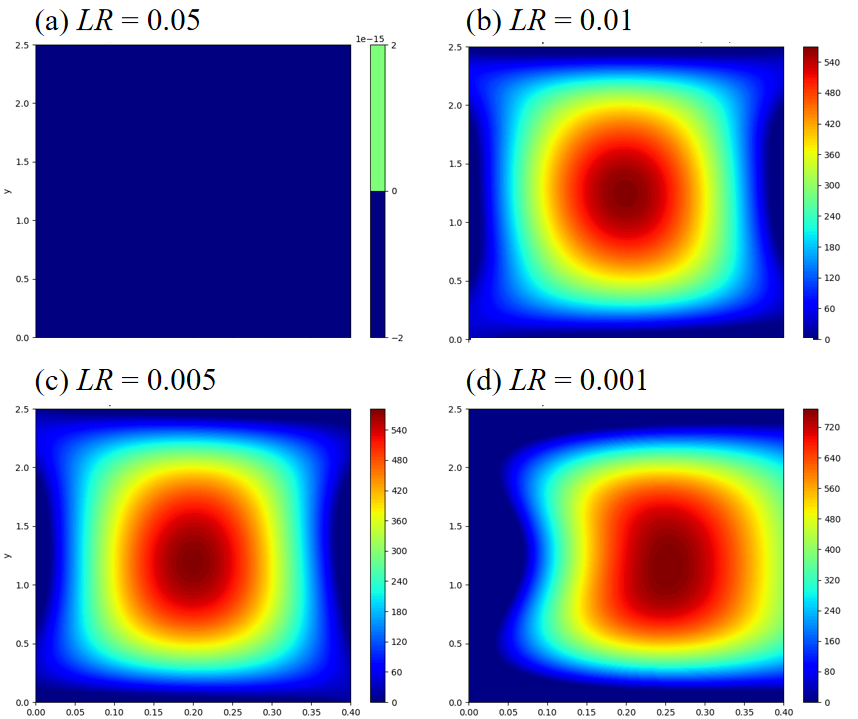}
	\caption{PINN results at $t=50\,\mathrm{s}$ for various learning rates ($LR$).}
	\label{F2}
\end{figure}

Figure \ref{F2} displays the temperature distribution predicted by the PINN at $t=50$ seconds with various learning rates. With a learning rate of 0.005, the model demonstrates excellent convergence and stability, producing a smooth temperature distribution with well-defined gradients, indicating that the model quickly reaches an optimal solution. With a learning rate of 0.01, convergence remains good, though some oscillations persist, still achieving fast convergence with acceptable stability. With a learning rate of 0.001, the model converges too slowly, and the temperature distribution becomes overly uniform, suggesting the model may be stuck in a local optimum, resulting in low training efficiency. With a learning rate of 0.05, the temperature distribution is highly irregular, indicating instability and potential gradient explosion.

\begin{figure}[htbp]
	\centering
	\includegraphics[scale=0.25]{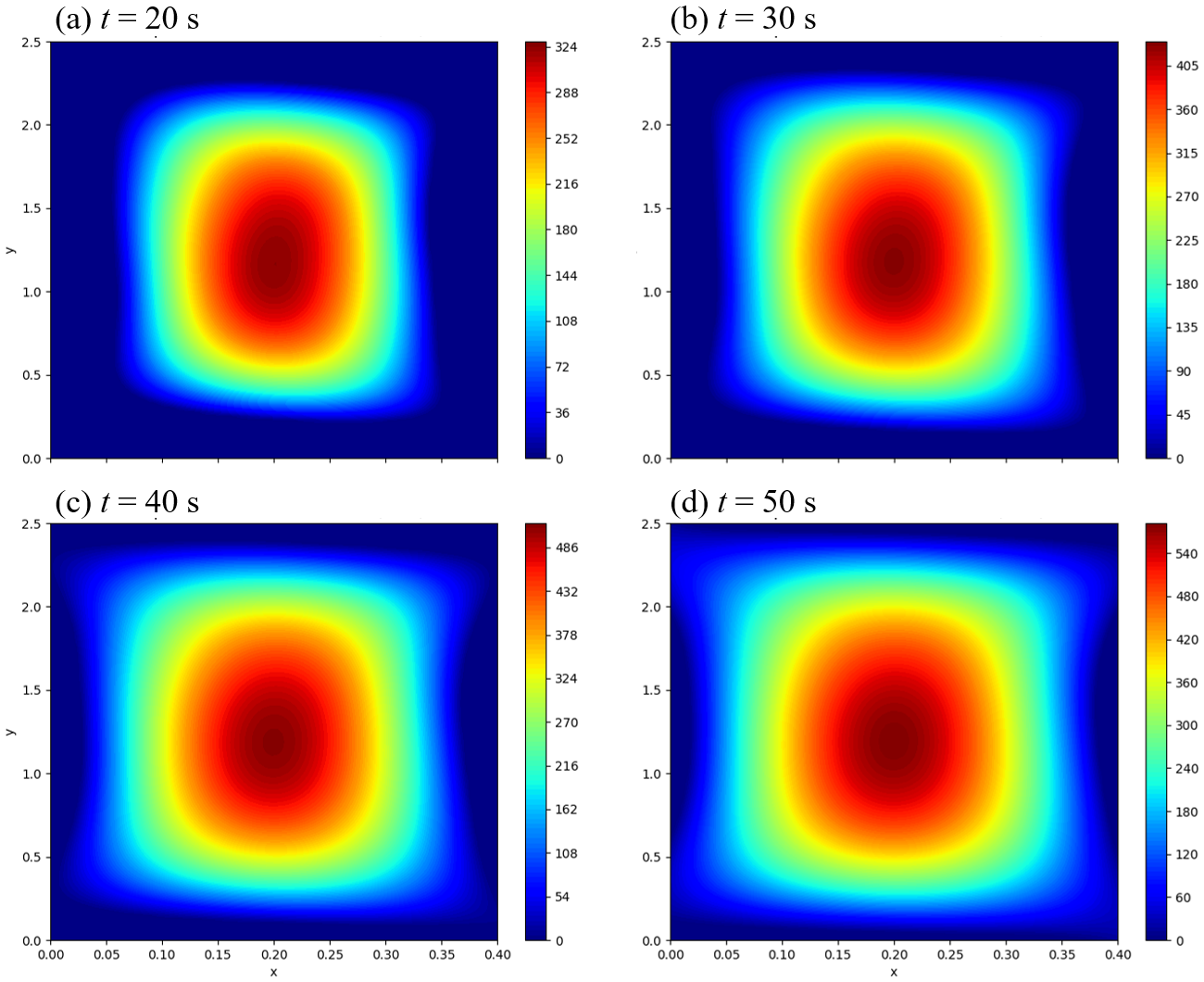}
	\caption{Results for a 0.005 learning rate at various times ($t$).}
	\label{F3}
\end{figure}

Figure \ref{F3} shows the temperature distribution at $t=20$, $t=30$, $t=40$, and $t=50$ seconds with a learning rate of 0.005. As time progresses, the red regions (indicating high-temperature areas) expand, demonstrating heat diffusion from the center. The evolving temperature gradient reflects the dynamic nature of the temperature field during heat conduction, with the color transition from red to blue signifying a decrease in temperature. These results validate the effectiveness of the PINN in simulating heat conduction problems, accurately capturing the temporal and spatial changes in temperature.

\section{Conclusion}
This study combines analytical solutions and neural network methods to solve the two-dimensional nonlinear transient heat conduction problem. The analytical solution is derived using pulse decomposition (Green’s function) and integral transform methods, both yielding the same general result. The integral transform method offers advantages in handling complex boundary conditions and reducing computational complexity. Additionally, the impact of various learning rates on the neural network model's performance was analyzed. A learning rate of 0.005 was found to provide the optimal balance between convergence speed and stability. The results highlight the importance of proper learning rate selection to ensure efficient neural network training. Overall, this research demonstrates the effectiveness of both analytical and neural network methods in solving heat conduction problems and provides valuable insights for future research and applications in engineering practice.




\printcredits
\section*{Declaration of competing interest}
The authors declared that they have no conflicts of interest to this work. We declare that we do not have any commercial or
associative interest that represents a conflict of interest in connection with the work submitted.

\section*{Acknowledgment}
This work is supported by the developing Project of Science and Technology of Jilin Province (20240402042GH). 

\section*{Data availability}
Data will be made available on request.

\appendix
\renewcommand{\theequation}{\thesection.\arabic{equation}}
\section{Calculation details of Green's function method}
\setcounter{equation}{0}
The second part of the boundary condition for a non - homogeneous problem\eqref{eqn-1a} with homogeneous boundary conditions is:
\begin{subequations}\label{eqn-2a}
	\begin{align}
		&\nabla^{2}G+\frac{\delta\left(\bm{r}-\bm{r}^{\prime}\right)}{k}\delta\left(t-\tau\right)=\frac{1}{\alpha}\frac{\partial G}{\partial t}\label{3.1.2a}\\
		B.C.\quad &k_{i}\frac{\partial G}{\partial n_{i}}+h_{i}G=0\quad t>\tau,\in S_{i}\\
		I.C.\quad &G=0,\quad t<\tau,\in R\label{3.1.2c}
	\end{align}
\end{subequations}
where $i=1,2,...,s$\\
with
\begin{equation}
	G=G\left(\bm{r},t|\bf{r}^{\prime},\tau\right),
\end{equation}
Derived from the principle of reciprocity, we obtain:
\begin{equation}
	G\left(\bm{r},t|\bm{r}^{\prime},\tau\right)=	G\left(\bm{r}^{\prime},-\tau|\bf{r},-\mathnormal{t}\right),\label{1b}
\end{equation}
Substituting Equation \eqref{1b} into \eqref{3.1.2a} yields:
\begin{equation}
	\nabla_{0}^{2}G+\frac{\delta\left(\bm{r}^{\prime}-\bm{r}\right)}{k}\delta\left(\tau-t\right)=-\frac{1}{\alpha}\frac{\partial G}{\partial \tau},\label{3.5}
\end{equation}
Here, the operator $\nabla_{0}^{2}$ denotes the Laplacian operator acting on functions dependent on the variable $\bm{r}^{\prime}$\\
Furthermore, if one replaces t with $\tau$ and $\bm{r}$ with $\bm{r}^{\prime}$ in formula$\left(\ref{3.1.1a}\right)$, we obtain:
\begin{equation}
	\nabla_{0}^{2}T+\frac{g\left(\bm{r}^{\prime},t\right)}{k}=\frac{1}{\alpha}\frac{\partial T\left(\bm{r}^{\prime},t\right)}{\partial \tau},\label{3.6}
\end{equation}
The combination of Equations \eqref{3.5} and \eqref{3.6} yields:
\begin{equation}
	G\nabla_{0}^{2}T-T\nabla_{0}^{2}G-\frac{1}{k}(g\left(\bm{r}^{\prime},\tau\right)G-T\delta\left(\bm{r}^{\prime}-\bm{r}\right)\delta\left(\tau-t\right))=\frac{1}{\alpha}\frac{\partial \left(GT\right)}{\partial \tau},
\end{equation}
Thus, we obtain:
\begin{equation}
	\int_{0}^{t^{*}}\mathrm{d}\tau\int_{R}(G\nabla_{0}^{2}T-T\nabla_{0}^{2}G)\mathrm{d}V^{\prime}+\frac{1}{k}\int_{0}^{t^{*}}gG\mathrm{d}V^{\prime}-\frac{T\left(\bm{r},t\right)}{k}=\frac{1}{\alpha}\int_{R}\left(GT\right)|_{0}^{t^{*}}\mathrm{d}V^{\prime},\label{4.9}
\end{equation}
where
\begin{equation}
		t^{*}=t+\epsilon,
\end{equation}
Here $\epsilon$ denotes an arbitrarily small quantity.
\\From Green's theorem:
\begin{equation}
	\int_{R}(G\nabla_{0}^{2}T-T\nabla_{0}^{2}G)\mathrm{d}V^{\prime}=\sum_{i=1}^{s}\int_{S_{i}}\left(G\frac{\partial T}{\partial n_{i}}-T\frac{\partial G}{\partial n_{i}}\right)\mathrm{d}S_{i},\label{4.11}
\end{equation}
Furthermore, we have:
\begin{equation}
	\left(GT\right)|_{\tau=0}^{\tau=t^{*}}=-\left(GT\right)|_{\tau=0}=-G|_{\tau=0}F\left(\bm{r}^{\prime}\right),\label{4.12}
\end{equation}
Substituting equation $\eqref{4.11}$, $\eqref{4.12}$ into equation $\eqref{4.9}$ and taking the limit as  $\tau\to 0$, we obtain:
\begin{equation}
	T\left(\bm{r},t\right)=\frac{k}{\alpha}I_{1}+I_{2}+kI_{3},\label{3.1.13}
\end{equation}
with the integrals:
\begin{subequations}\label{3.1.14}
	\begin{align}
		&I_{1}=\int_{R}G|_{\tau=0}F\left(\bm{r}\right)\mathrm{d}V^{\prime},\\
		&I_{2}=\int_{0}^{t}\mathrm{d}\tau\int_{R}G\left(\bm{r},t|\bm{r}^{\prime},\tau\right)g\left(\bm{r}^{\prime},t\right)\mathrm{d}V^{\prime},\\
		&I_{3}=\int_{0}^{t}\mathrm{d}\tau\sum_{i=1}^{s}\int_{S_{i}}\left(G\frac{\partial T}{\partial n_{i}}-T\frac{\partial G}{\partial n_{i}}\right)\mathrm{d}S_{i},
	\end{align}
\end{subequations}
and
\begin{equation}
	G|_{\tau=0}=G\left(\bm{r},t|\bm{r}^{\prime},0\right),
\end{equation}
Let $G|_{S_{i}}$ denote the value of the Green's function on the boundary.\\
Use the boundary between problem \eqref{eqn-1a} and problem \eqref{eqn-2a} there:
\begin{equation}
	G\frac{\partial T}{\partial n_{i}}-T\frac{\partial G}{\partial n_{i}}=\frac{1}{k_{i}}G|_{S_{i}}f_{i}\left(\bm{r}^{\prime},\tau\right),\label{3.1.16}
\end{equation}
From equations \eqref{3.1.13}, \eqref{3.1.14} and \eqref{3.1.16} we obtain:
\begin{equation}
	T\left(\bm{r},t\right)=\frac{k}{\alpha}I_{1}+I_{2}+kI_{3}^{\prime},\label{3.1.17}
\end{equation}
where 
\begin{equation}
	I_{3}^{\prime}=\int_{0}^{t}\mathrm{d}\tau\sum_{i=1}^{s}\int_{S_{i}}\frac{1}{k_{i}}G|_{r^{\prime}=r_{i}}f_{i}\left(\bm{r}^{\prime},\tau\right)\mathrm{d}S_{i},
\end{equation}
According to the principle of conservation of energy, the heat absorbed due to the temperature increase in the differential volume $\mathrm{d}V$ at position $\bm{r}$ during the interval from $t=\tau^{-}$ to $t=\tau$ equals the heat generated by the internal heat source. Hence, we obtain:
\begin{equation}
	\frac{k}{\alpha}\Delta G\mathrm{d}V=\delta\left(\bm{r}-\bm{r}^{\prime}\right)\delta\left(t-\tau\right)\mathrm{d}V\mathrm{d}\tau,\label{3.1.19}
\end{equation}
where
\begin{equation}
	\Delta G=G|_{t=\tau}-G|_{t=\tau^{-}},
\end{equation}
Combining equation \eqref{3.1.19} with equation \eqref{3.1.2c} then yields:
\begin{equation}
	G|_{t=\tau}=\frac{\alpha}{k}\delta\left(\bm{r}-\bm{r}^{\prime}\right),
\end{equation}
Thus, question \eqref{eqn-2a} is equivalent to:
\begin{subequations}\label{3.1.22}
	\begin{align}
		&\nabla^{2}G\left(\bm{r},t|\bm{r}^{\prime},\tau\right)=\frac{1}{\alpha}\frac{\partial G}{\partial t}\label{3.1.22a}\\
		B.C.\quad &k_{i}\frac{\partial G}{\partial n_{i}}+h_{i}G=0\quad t>\tau,\in S_{i}\\
		I.C.\quad &G=\frac{\alpha}{k}\delta\left(\bm{r}-\bm{r}^{\prime}\right),\quad t=\tau,\in R\label{3.1.22c}
	\end{align}
\end{subequations}
where $i=1,2,...,s$, and $t>\tau,\in R$\\
Let
\begin{equation}
	G=\mathfrak{A}\left(t\right)\psi\left(\bm{r}\right)\label{3.1.23}
\end{equation}
By separating the spatiotemporal variables, we obtain the following intrinsic problem:
\begin{subequations}\label{3.1.24}
	\begin{align}
		&\nabla^{2}\psi\left(\bm{r}\right)+\lambda^{2}\psi\left(\bm{r}\right)\label{3.1.24a}\\
		B.C.\quad &k_{i}\frac{\partial \psi}{\partial n_{i}}+h_{i}\psi=0\quad\in S_{i}
	\end{align}
\end{subequations}
and substituting equation \eqref{3.1.23} into equation \eqref{3.1.22a} yields:
\begin{equation}
	\mathfrak{A}\left(t\right)=e^{-\alpha\lambda_{m}^{2}t},\label{3.1.25}
\end{equation}
Returning to the one-dimensional eigenproblem, we obtain:
\begin{subequations}\label{3.1.26}
	\begin{align}
		&X^{\prime\prime}\left(x\right)+\beta^{2}X\left(x\right)=0, \quad0<x<L\\
		&-k_{1}X^{\prime}+h_{1}X=0,\quad x=0\label{3.1.26b}\\
		&k_{2}X^{\prime}+h_{2}X=0,\quad x=L\label{3.1.26c}
	\end{align}
\end{subequations}
The general solution is given by:
\begin{equation}
	X\left(x\right)=C_{1}\cos\beta x+C_{2}\sin\beta x,\label{3.1.27}
\end{equation}
Substituting equation \eqref{3.1.27} into equations \eqref{3.1.26b} and \eqref{3.1.26c} yields:
\begin{equation}\label{3.1.28}
	\bm{M}\left(\begin{matrix}
		C_{1}\\
		C_{2}
	\end{matrix}\right)=0,
\end{equation}
where
\begin{equation}
	\bm{M}=\left(\begin{matrix}
		\frac{h_{1}}{k_{1}}\quad-\beta\\
		\frac{h_{2}}{k_{2}}\cos\beta L-\beta\sin\beta L\quad\frac{h_{2}}{k_{2}}\sin\beta L+\beta\cos\beta L
	\end{matrix}\right),
\end{equation}
Thus, we obtain:
\begin{equation}
	\left|\bm{M}\right|=0,
\end{equation}
Let
\begin{equation}
	H_{i}=\frac{h_{i}}{k_{i}},\quad i=1,2
\end{equation}
Thus, we obtain:
\begin{equation}
	\tan\beta L=\frac{H_{1}+H_{2}}{\beta^{2}-H_{1}H_{2}}\beta,\label{3.1.34}
\end{equation}
The eigenvalues $\beta=\beta_{m}$, $\left(m=1,2,...\right)$—of which there are infinitely many—are the roots of equation \eqref{3.1.34}. From equations \eqref{3.1.27} and \eqref{3.1.28} corresponding eigenfunction is given by:
\begin{equation}
	X\left(\beta_{m},x\right)=\beta_{m}\cos\beta_{m}x+H_{1}\sin\beta_{m}x,\label{3.1.35}
\end{equation}
where $m=1,2,...$\\
It satisfies the orthogonality condition:
\begin{equation}
	\int_{0}^{L}X\left(\beta_{m},x\right)X\left(\beta_{n},x\right)\mathrm{d}x=\left\{
	\begin{aligned}
		&0,\quad &m\neq n\\
		&N\left(\beta_{m}\right),\quad &m=n
	\end{aligned}
	\right.
\end{equation}
where
\begin{equation}
	\begin{split}
		N\left(\beta_{m}\right)&=\int_{0}^{L}X^{2}\left(\beta_{m},x\right)\mathrm{d}x\\&=\frac{\beta_{m}^{2}+H_{m}^{2}}{2}\left(L+\frac{H_{2}}{\beta_{m}^{2}+H_{m}^{2}}\right)+\frac{H_{1}}{2},
	\end{split}
\end{equation}
Thus, for one-dimensional homogeneous problems:
\begin{subequations}\label{3.1.37}
	\begin{align}
		\frac{\partial^{2}\theta\left(x,t\right)}{\partial x^{2}}=\frac{1}{\alpha}\frac{\partial\theta\left(x,t\right)}{\partial t},\quad0<x<L,\quad t>0\label{3.1.37a}\\
		B.C.\quad-k_{1}\frac{\partial\theta}{\partial x}+h_{1}\theta=0,\quad x=0,\quad t>0 \\
		k_{2}\frac{\partial\theta}{\partial x}+h_{2}\theta=0,\quad x=L,\quad t>0 \\
		I.C.\quad\quad\quad\theta=I\left(x\right),\quad t=0,\quad0\leq x\leq L
	\end{align}
\end{subequations}
Let
\begin{equation}
	\theta\left(x,t\right)=X\left(x\right)\mathfrak{A}\left(t\right),\label{3.1.38}
\end{equation}
Substituting equation \eqref{3.1.38} into equation \eqref{3.1.37a}, we obtain:
\begin{equation}
	\frac{X^{\prime\prime}_{\left(x\right)}}{X\left(x\right)}=\frac{\mathfrak{A}^{\prime}_{\left(t\right)}}{\alpha\mathfrak{A}\left(t\right)},
\end{equation}
The expression on the left-hand side depends solely on the variable $x$, while the expression on the right-hand side depends solely on the variable $t$, Since these expressions are equal for all $x$ and $t$, they must both equal a constant, denoted as $-\beta^{2}$. Thus, we have:
\begin{equation}
	\mathfrak{A}^{\prime}\left(t\right)+\alpha\beta^{2}\mathfrak{A}\left(t\right)=0,
\end{equation}
Thus, we have:
\begin{equation}
	\mathfrak{A}\left(t\right)=e^{-\alpha\beta^{2}t},
\end{equation}
This is consistent with equation \eqref{3.1.25}. Moreover,  $X\left(x\right)$ satisfies problem \eqref{3.1.26}, and the corresponding eigenfunction system $\left\{X\left(\beta_{m},x\right)\right\}$ is given by equation \eqref{3.1.35}, Consequently, the initial state condition $I\left(x\right)$ can be expressed as:
\begin{equation}
	I\left(x\right)=\sum_{m=1}^{\infty}C_{m}X\left(\beta_{m},x\right),\quad0\leq x\leq L\label{3.1.42}
\end{equation}
where $C_{m}$ is called the expansion coefficient.\\
Consider equation \eqref{3.1.42}  as applied to each individual temperature field $C_ {m}X\left (\beta_ {m}, x \right)$, which is generated by the term $C_ {m}X\left(\beta_{m},x\right)\mathfrak{A}_{m}\left(t\right)$, . Therefore, if the contributions of all such terms to the temperature field are superimposed, the solution of problem \eqref{3.1.37} is given by:
\begin{equation}
	\theta\left(x,t\right)=\sum_{m=1}^{\infty}C_{m}X_{m}\mathfrak{A}_{m},\label{3.1.43}
\end{equation}
where
\begin{equation}
	\mathfrak{A}_{m}\left(t\right)=e^{-\alpha\lambda_{m}^{2}t},\label{3.1.44}
\end{equation}
The general solution of problem \eqref{3.1.22}, as compared to that of problem \eqref{3.1.37} is:
\begin{equation}
	G=\sum_{m=1}^{\infty}C_{m}\mathfrak{A}_{m}\psi\left(\lambda_{m},\bm{r}\right),\label{3.1.45}
\end{equation}
Substituting equation \eqref{3.1.44} into equation
\eqref{3.1.45} yields:
\begin{equation}
	G=\sum_{m=1}^{\infty}C_{m}e^{-\alpha\lambda_{m}^{2}t}\psi\left(\lambda_{m},\bm{r}\right),\label{3.1.46}
\end{equation}
Let
\begin{equation}
	t=\tau,\label{3.1.47}
\end{equation}
Substituting equation\eqref{3.1.47} and \eqref{3.1.46} into equation \eqref{3.1.22c} yields:
\begin{equation}
	\frac{\alpha}{k}\delta\left(\bm{r}-\bm{r}^{\prime}\right)=\sum_{m=1}^{\infty}C_{m}e^{-\alpha\lambda_{m}^{2}\tau}\psi\left(\lambda_{m},\bm{r}\right),\label{3.1.48}
\end{equation}
Exploiting the orthogonality of $\psi\left(\lambda_{m},\bm{r}\right)$, we apply the operator $\int_{R}\psi\left(\lambda_{m},\bm{r}\right)\mathrm{d}V$ to both sides of equation \eqref{3.1.48}, thereby obtaining:
\begin{equation}
	\int_{R}\frac{\alpha}{k}\delta\left(\bm{r}-\bm{r}^{\prime}\right)\psi\left(\lambda_{m},\bm{r}\right)\mathrm{d}V=C_{m}e^{-\alpha\lambda_{m}^{2}\tau}N\left(\lambda_{m}\right),\label{3.1.49}
\end{equation}
And
\begin{equation}
	\psi\left(\lambda_{m},\bm{r}^{\prime}\right)=\int_{R}\delta\left(\bm{r}-\bm{r}^{\prime}\right)\psi\left(\lambda_{m},\bm{r}\right)\mathrm{d}V,\label{3.1.50}
\end{equation}
Substituting equation \eqref{3.1.50} into equation \eqref{3.1.49} yields:
\begin{equation}
	C_{m}=\frac{\alpha e^{\alpha\lambda_{m}^{2}\tau}}{kN\left(\lambda_{m}\right)}\psi\left(\lambda_{m},\bm{r}^{\prime}\right),\label{3.1.51}
\end{equation}
Substituting equation \eqref{3.1.51} into equation \eqref{3.1.46} yields:
\begin{equation}\label{3.1.a}
G\left(\bm{r},t|\bm{r}^{\prime},\tau\right)=\frac{\alpha}{k}\sum_{m=1}^{\infty}\frac{\psi\left(\lambda_{m},\bm{r}\right)\psi\left(\lambda_{m},\bm{r}^{\prime}\right)}{N\left(\lambda_{m}\right)}e^{-\alpha\lambda_{m}^{2}\left(t-\tau\right)},
\end{equation}
Substituting equation \eqref{3.1.a} into equation \eqref{3.1.17} yields:
\begin{equation}
	T\left(\bm{r},t\right)=I_{1}^{*}+\frac{\alpha}{k}I_{2}^{*}+\alpha I_{3}^{*},
\end{equation}
where
\begin{subequations}
	\begin{align}
		&I_{1}^{*}=\int_{R}\sum_{m=1}^{\infty}\frac{e^{-\alpha\lambda_{m}^{2}t}}{N\left(\lambda_{m}\right)}\psi\left(\lambda_{m},\bm{r}^{\prime}\right)F\left(\bm{r}^{\prime}\right)\mathrm{d}V^{\prime},\\
		&I_{2}^{*}=\int_{0}^{t}\mathrm{d}\tau\int_{R}\sum_{m=1}^{\infty}\frac{\psi\left(\lambda_{m},\bm{r}\right)\psi\left(\lambda_{m},\bm{r}^{\prime}\right)}{N\left(\lambda_{m}\right)}e^{-\alpha\lambda_{m}^{2}\left(t-\tau\right)}g\left(\bm{r}^{\prime},\tau\right)\mathrm{d}V^{\prime},\\
		&I_{3}^{*}=\int_{0}^{t}\mathrm{d}\tau\sum_{i=1}^{s}\int_{S_{i}}\sum_{m=1}^{\infty}\frac{\psi\left(\lambda_{m},\bm{r}\right)\psi\left(\lambda_{m},\bm{r}^{\prime}\right)}{k_{i}N\left(\lambda_{m}\right)}e^{-\alpha\lambda_{m}^{2}\left(t-\tau\right)}f_{i}\left(\bm{r}^{\prime},\tau\right)\mathrm{d}S_{i},
	\end{align}
\end{subequations}
After tidying up and replacing $\tau$ with $t^{\prime}$, we can obtain equation \eqref{3.1.56}.
\section{Calculation details of Integral transformation method}
\setcounter{equation}{0}
The homogeneous component of problem \eqref{eqn-1a} is expressed as:
\begin{subequations}
	\begin{align}
		&\nabla^{2}T^{*}\left(\bf{r},\mathnormal{t}\right)=\frac{1}{\alpha}\frac{\partial T^{*}\left(\bf{r},\mathnormal{t}\right)}{\partial t}\label{1a},\in R\\
		B.C.\quad &k_{i}\frac{\partial T^{*}}{\partial n_{i}}+h_{i}T^{*}=0,\quad\in S_{i}\\\label{3.2.68b}
		I.C.\quad &T^{*}=F\left(\bf{r}\right),\quad\in R
	\end{align}
\end{subequations}
where $t>0,\in R$ and $i=1,2,...,s$\\
Let
\begin{equation}
	T^{*}\left(\bf{r},\mathnormal{t}\right)=\psi\left(\bm{r}\right)\mathfrak{A}\left(t\right),
\end{equation}
Furthermore, the same intrinsic problem as in Equation \eqref{3.1.24} can be obtained. In the same way as for the one - dimensional eigenproblem \eqref{3.1.26}, the eigenfunctions form a complete orthogonal system:
\begin{equation}\label{3.2.70}
	\int_{0}^{L}\psi\left(\lambda_{m},\bm{r}\right)\psi\left(\lambda_{n},\bm{r}\right)\mathrm{d}V=\left\{
	\begin{aligned}
		&0,\quad &m\neq n\\
		&N\left(\lambda_{m}\right),\quad &m=n
	\end{aligned}
	\right.
\end{equation}
Subsequently,  the temperature field $T\left(\bm{r},t\right)$ in problem \eqref{eqn-1a} can be expanded as:
\begin{equation}
	T\left(\bm{r},t\right)=\sum_{m=1}^{\infty}C_{m}\psi\left(\lambda_{m},\bm{r}\right),\label{3.2.71}
\end{equation}
Furthermore, applying the operator 
$\int_{R}\psi\left(\lambda_{m},\bm{r}\right)\mathrm{d}V$  to both sides of equation \eqref{3.2.71}, and using the orthogonal relation of equation \eqref{3.2.70}, we obtain:
\begin{equation}\label{3.2.72}
	C_{m}=\frac{1}{N\left(\lambda_{m}\right)}\int_{R}\psi\left(\lambda_{m},\bm{r}\right)T\left(\bm{r},t\right)\mathrm{d}V,
\end{equation}
Substituting equation \eqref{3.2.72} into \eqref{3.2.71} yields:
\begin{equation}
	T\left(\bm{r},t\right)=\sum_{m=1}^{\infty}\frac{\psi\left(\lambda_{m},\bm{r}\right)}{N\left(\lambda_{m}\right)}\int_{R}\psi\left(\lambda_{m},\bm{r}^{\prime}\right)T\left(\lambda_{m},\bm{r}^{\prime}\right)\mathrm{d}V^{\prime},
\end{equation}
The contravariant formula is therefore defined as follows:
\begin{equation}
	T\left(\bm{r},t\right)=\sum_{m=1}^{\infty}\frac{\psi\left(\lambda_{m},\bm{r}\right)}{N\left(\lambda_{m}\right)}\overline{T}\left(\lambda_{m},t\right),\label{3.2.740}
\end{equation}
The integral transformation is formally defined by:
\begin{equation}
	\overline{T}\left(\lambda_{m},t\right)=\int_{R}\psi\left(\lambda_{m},\bm{r}^{\prime}\right)T\left(\lambda_{m},\bm{r}^{\prime}\right)\mathrm{d}V^{\prime},\label{3.2.75}
\end{equation}
Furthermore, applying the operator $\int_{R}\psi\left(\lambda_{m},\bm{r}\right)\mathrm{d}V$ to both sides of equation \eqref{3.1.1a}, we obtain:
\begin{equation}
	\int_{R}\psi\left(\lambda_{m},\bm{r}\right)\nabla^{2}T\left(\bm{r},t\right)\mathrm{d}V+\frac{1}{k}\overline{g}\left(\lambda_{m},t\right)=\frac{1}{\alpha}\frac{\partial\overline{T}\left(\lambda_{m},t\right)}{\mathrm{d}t},\label{3.2.76}
\end{equation}
where $\overline{g}\left(\lambda_{m},t\right)$ and $\overline{T}\left(\lambda_{m},t\right)$ satisfy equation \eqref{3.2.75}, and
\begin{equation}
	\int_{R}\psi\left(\lambda_{m},\bm{r}\right)\nabla^{2}T\left(\bm{r},t\right)\mathrm{d}V=I_{a_0}+I_{b_0},\label{3.2.77}
\end{equation}
with the integrals:
\begin{equation}
I_{a_0}=\int_{R}T\nabla^{2}\psi\left(\lambda_{m},\bm{r}\right)\mathrm{d}V,\label{3.2.78}
\end{equation}
\begin{equation}
I_{b_0}=\sum_{i=1}^{s}\int_{S_{i}}\left(\psi\frac{\partial T}{\partial n_{i}}-T\frac{\partial\psi}{\partial n_{i}}\right)\mathrm{d}S_{i},\label{3.2.79}
\end{equation}
By combining equation \eqref{3.1.24a} with $\lambda_{m}$, the following result can be derived:
\begin{equation}
	\nabla^{2}\psi\left(\lambda_{m},\bm{r}\right)=-\lambda_{m}\psi\left(\lambda_{m},\bm{r}\right),\label{3.2.80}
\end{equation}
Substituting Equation \eqref{3.2.75} and equation \eqref{3.2.78} into equation \eqref{3.2.80}, we obtain:
\begin{equation}\label{3.2.81}
	I_{a_0}=-\lambda_{m}^{2}\int_{R}T\psi\left(\lambda_{m},\bm{r}\right)\mathrm{d}V=-\lambda_{m}^{2}\overline{T}\left(\lambda_{m},t\right),
\end{equation}
Moreover, $T\left(\bm{r},t\right)$ and $\psi\left(\lambda_{m},\bm{r}\right)$ satisfy equation \eqref{3.1.1b} and equation \eqref{3.2.68b}, respectively. Thus, we have:
\begin{equation}
	k_{i}\frac{\partial T}{\partial n_{i}}+h_{i}T=f_{i}\left(\bm{r},t\right)\label{3.2.82},
\end{equation}
and
\begin{equation}
	k_{i}\frac{\partial\psi}{\partial n_{i}}+h_{i}\psi=0,\label{3.2.83}
\end{equation}
In addition, by multiplying equation \eqref{3.2.82} by the function $\psi\left(\lambda_{m},\bm{r}\right)$ and equation \eqref{3.2.83} by the function $T\left(\bm{r},t\right)$, and subsequently subtracting the two equations, we obtain the following result:
\begin{equation}
	\psi\frac{\partial T}{\partial n_{i}}-T\frac{\partial\psi}{\partial n_{i}}=\frac{\psi\left(\lambda_{m},\bm{r}\right)}{k_{i}}f_{i}\left(\bm{r},t\right),\label{3.2.84}
\end{equation}
where $i=1,2,...,s$\\
By substituting Equation \eqref{3.2.84} into equation \eqref{3.2.79}, we obtain:
\begin{equation}\label{3.2.85}
	I_{b_0}=\sum_{i=1}^{s}\int_{S_{i}}\frac{\psi\left(\lambda_{m},\bm{r}\right)}{k_{i}}f_{i}\left(\bm{r},t\right)\mathrm{d}S_{i},
\end{equation}
By substituting equations \eqref{3.2.81}, \eqref{3.2.85}, and \eqref{3.2.77} into equation \eqref{3.2.76}, we obtain:
\begin{equation}
	\frac{\mathrm{d}}{\mathrm{d}t}\overline{T}\left(\lambda_{m},t\right)=\alpha\left(-\lambda_{m}^{2}\overline{T}\left(\lambda_{m},t\right)+\frac{1}{k}\overline{g}\left(\lambda_{m},t\right)+\sum_{i=1}^{s}\int_{S_{i}}\frac{\psi\left(\lambda_{m},\bm{r}\right)}{k_{i}}f_{i}\left(\bm{r},t\right)\mathrm{d}S_{i}\right),\label{3.2.86}
\end{equation}
We apply the operator  $\int_{R}\psi\left(\lambda_{m},\bm{r}\right)$ to both sides of Equation \eqref{3.1.1c}, yielding:
\begin{equation}
	\int_{R}\psi\left(\lambda_{m},\bm{r}\right)T\left(\lambda_{m},t\right)\mathrm{d}V=\int_{R}\psi\left(\lambda_{m},\bm{r}\right)F\left(\bm{r}\right)\mathrm{d}V,\label{3.2.87}
\end{equation}
By substituting equation \eqref{3.2.75} into equation \eqref{3.2.87}, we obtain:
\begin{equation}
	\overline{T}\left(\lambda_{m},t\right)=\overline{F}\left(\lambda_{m}\right),\label{3.2.88}
\end{equation}
where $\overline{F}\left(\lambda_{m}\right)$ satisfies the equation \eqref{3.2.75}.\\
From equations \eqref{3.2.86} and \eqref{3.2.88}, we obtain:
\begin{subequations}\label{3.2.90}
	\begin{align}
		&\frac{\mathrm{d}\overline{T}\left(\lambda_{m},t\right)}{\mathrm{d}t}+\alpha\lambda_{m}^{2}\overline{T}\left(\lambda_{m},t\right)=\alpha\left(\frac{1}{k}\overline{g}\left(\lambda_{m},t\right)+\sum_{i=1}^{s}\int_{S_{i}}\frac{\psi\left(\lambda_{m},\bm{r}\right)}{k_{i}}f_{i}\left(\bm{r},t\right)\mathrm{d}S_{i}\right),\in R\label{3.2.90a}\\
		I.C.\quad &\overline{T}\left(\lambda_{m},t\right)=\overline{F}\left(\lambda_{m}\right),\quad t=0,\in R \label{3.2.90b}
	\end{align}
\end{subequations}
Therefore, the solution to problem \eqref{3.2.90} can be expressed as follows:
\begin{equation}
	\overline{T}\left(\lambda_{m},t\right)=\left(\int_{0}^{t}e^{\alpha\lambda_{m}^{2}t^{\prime}}\alpha\left(\frac{1}{k}\overline{g}\left(\lambda_{m},t^{\prime}\right)+\sum_{i=1}^{s}\int_{S_{i}}\frac{\psi\left(\lambda_{m},\bm{r}\right)}{k_{i}}f_{i}\left(\bm{r},t^{\prime}\right)\mathrm{d}S_{i}\right)\mathrm{d}t^{\prime}+\overline{F}\left(\lambda_{m}\right)\right)e^{-\alpha\lambda_{m}^{2}t},\label{3.2.910}
\end{equation}
By substituting equation \eqref{3.2.910} into equation \eqref{3.2.740}, we can obtain equation \eqref{3.2.74}.
\bibliographystyle{cas-model2-names}

\bibliography{cas-refs}

\begin{thebibliography}{27}
\expandafter\ifx\csname natexlab\endcsname\relax\def\natexlab#1{#1}\fi
\providecommand{\url}[1]{\texttt{#1}}
\providecommand{\href}[2]{#2}
\providecommand{\path}[1]{#1}
\providecommand{\DOIprefix}{doi:}
\providecommand{\ArXivprefix}{arXiv:}
\providecommand{\URLprefix}{URL: }
\providecommand{\Pubmedprefix}{pmid:}
\providecommand{\doi}[1]{\href{http://dx.doi.org/#1}{\path{#1}}}
\providecommand{\Pubmed}[1]{\href{pmid:#1}{\path{#1}}}
\providecommand{\bibinfo}[2]{#2}
\ifx\xfnm\relax \def\xfnm[#1]{\unskip,\space#1}\fi
\bibitem[{Woodbury et~al.(2023)Woodbury, Najafi, De~Monte and Beck}]{ref9}
\bibinfo{author}{Woodbury, K.A.}, \bibinfo{author}{Najafi, H.}, \bibinfo{author}{De~Monte, F.}, \bibinfo{author}{Beck, J.V.}, \bibinfo{year}{2023}.
\newblock \bibinfo{title}{Inverse heat conduction: Ill-Posed problems}.
\newblock \bibinfo{publisher}{John Wiley \& Sons}.
\bibitem[{Alifanov(2012)}]{ref11}
\bibinfo{author}{Alifanov, O.M.}, \bibinfo{year}{2012}.
\newblock \bibinfo{title}{Inverse heat transfer problems}.
\newblock \bibinfo{publisher}{Springer Science \& Business Media}.
\bibitem[{Xia et~al.(2025)Xia, Yang, Zheng and Wang}]{ref30}
\bibinfo{author}{Xia, Y.}, \bibinfo{author}{Yang, Y.}, \bibinfo{author}{Zheng, H.}, \bibinfo{author}{Wang, S.}, \bibinfo{year}{2025}.
\newblock \bibinfo{title}{Modelling of thermo-mechanical coupling effects in rock masses using an enriched nodal-based continuous-discontinuous deformation analysis method}.
\newblock \bibinfo{journal}{COMPUTER METHODS IN APPLIED MECHANICS AND ENGINEERING} \bibinfo{volume}{433}.
\newblock \DOIprefix\doi{10.1016/j.cma.2024.117543}.
\bibitem[{Daryabeigi(1999)}]{ref7}
\bibinfo{author}{Daryabeigi, K.}, \bibinfo{year}{1999}.
\newblock \bibinfo{title}{Effective thermal conductivity of high temperature insulations for reusable launch vehicles}. volume~\bibinfo{volume}{98}.
\newblock \bibinfo{publisher}{NASA}.
\bibitem[{Cohen et~al.(2019)Cohen, Padan, Shuraki and Khoptiar}]{ref8}
\bibinfo{author}{Cohen, Y.}, \bibinfo{author}{Padan, R.}, \bibinfo{author}{Shuraki, G.}, \bibinfo{author}{Khoptiar, Y.}, \bibinfo{year}{2019}.
\newblock \bibinfo{title}{Development of an sma based device for the thermal management of electronic systems}, in: \bibinfo{booktitle}{International Conference on Shape Memory and Superelastic Technologies 2019, SMST 2019}, \bibinfo{organization}{ASM International}. p.~\bibinfo{pages}{80}.
\bibitem[{Jiang et~al.(2020)Jiang, Liu, Yang and Gao}]{ref31}
\bibinfo{author}{Jiang, G.}, \bibinfo{author}{Liu, H.}, \bibinfo{author}{Yang, K.}, \bibinfo{author}{Gao, X.}, \bibinfo{year}{2020}.
\newblock \bibinfo{title}{A fast reduced-order model for radial integral boundary element method based on proper orthogonal decomposition in nonlinear transient heat conduction problems}.
\newblock \bibinfo{journal}{COMPUTER METHODS IN APPLIED MECHANICS AND ENGINEERING} \bibinfo{volume}{368}.
\newblock \DOIprefix\doi{10.1016/j.cma.2020.113190}.
\bibitem[{Rolfes(1991)}]{ref13}
\bibinfo{author}{Rolfes, R.}, \bibinfo{year}{1991}.
\newblock \bibinfo{title}{Higher order theory and finite element for heat conduction in composites.}, in: \bibinfo{booktitle}{Proc. 7th International Conf. on Numerical Methods in Thermal Problems.}, pp. \bibinfo{pages}{880--889}.
\bibitem[{Rolfes(1992)}]{ref14}
\bibinfo{author}{Rolfes, R.}, \bibinfo{year}{1992}.
\newblock \bibinfo{title}{Nonlinear transient thermal analysis of composite structures.}, in: \bibinfo{booktitle}{Tagungsband:" Numerical Methods in Engineering'92" Proc. of the First European Conference on Numerical Methodsin Eng.}, \bibinfo{organization}{Elsevier, Amsterdam-London-New York-Tokyo}. pp. \bibinfo{pages}{653--659}.
\bibitem[{Tamma and Yurko(1988)}]{ref15}
\bibinfo{author}{Tamma, K.K.}, \bibinfo{author}{Yurko, A.A.}, \bibinfo{year}{1988}.
\newblock \bibinfo{title}{A unified finite element modeling/analysis approach for thermal-structural response in layered composites}.
\newblock \bibinfo{journal}{Computers \& structures} \bibinfo{volume}{29}, \bibinfo{pages}{743--754}.
\bibitem[{Shyy(2002)}]{ref16}
\bibinfo{author}{Shyy, W.}, \bibinfo{year}{2002}.
\newblock \bibinfo{title}{Multi-scale computational heat transfer with moving solidification boundaries}.
\newblock \bibinfo{journal}{International journal of heat and fluid flow} \bibinfo{volume}{23}, \bibinfo{pages}{278--287}.
\bibitem[{Wang et~al.(2005)Wang, Qin and Kang}]{ref17}
\bibinfo{author}{Wang, H.}, \bibinfo{author}{Qin, Q.H.}, \bibinfo{author}{Kang, Y.L.}, \bibinfo{year}{2005}.
\newblock \bibinfo{title}{A new meshless method for steady-state heat conduction problems in anisotropic and inhomogeneous media}.
\newblock \bibinfo{journal}{Archive of Applied Mechanics} \bibinfo{volume}{74}, \bibinfo{pages}{563}.
\bibitem[{Onate et~al.(2022)Onate, de~Pouplana and Zarate}]{ref29}
\bibinfo{author}{Onate, E.}, \bibinfo{author}{de~Pouplana, I.}, \bibinfo{author}{Zarate, F.}, \bibinfo{year}{2022}.
\newblock \bibinfo{title}{Explicit time integration scheme with large time steps for first order transient problems using finite increment calculus}.
\newblock \bibinfo{journal}{COMPUTER METHODS IN APPLIED MECHANICS AND ENGINEERING} \bibinfo{volume}{402}.
\bibitem[{Fr{\k{a}}ckowiak and Cia{\l}kowski(2014)}]{ref19}
\bibinfo{author}{Fr{\k{a}}ckowiak, A.}, \bibinfo{author}{Cia{\l}kowski, M.}, \bibinfo{year}{2014}.
\newblock \bibinfo{title}{Green’s functions in steady-state heat conduction}.
\newblock \bibinfo{journal}{Encyclopedia of thermal stresses. Springer, Dordrecht. https://doi. org/10.1007/978-94-007-2739-7} .
\bibitem[{K{\v{r}}{\'\i}{\v{z}}ek and Liu(1998)}]{ref20}
\bibinfo{author}{K{\v{r}}{\'\i}{\v{z}}ek, M.}, \bibinfo{author}{Liu, L.}, \bibinfo{year}{1998}.
\newblock \bibinfo{title}{Finite element approximation of a nonlinear heat conduction problem in anisotropic media}.
\newblock \bibinfo{journal}{Computer methods in applied mechanics and engineering} \bibinfo{volume}{157}, \bibinfo{pages}{387--397}.
\bibitem[{Ramos and Rossi(2019)}]{ref32}
\bibinfo{author}{Ramos, G.R.}, \bibinfo{author}{Rossi, R.}, \bibinfo{year}{2019}.
\newblock \bibinfo{title}{A novel computational multiscale approach to model thermochemical coupled problems in heterogeneous solids: Application to the determination of the “state of cure” in filled elastomers}.
\newblock \bibinfo{journal}{Computer Methods in Applied Mechanics and Engineering} \bibinfo{volume}{351}, \bibinfo{pages}{694--717}.
\bibitem[{Itagi(2007)}]{ref21}
\bibinfo{author}{Itagi, A.}, \bibinfo{year}{2007}.
\newblock \bibinfo{title}{Finite volume method for the fourier heat conduction in layered media with a moving volume heat source}.
\newblock \bibinfo{journal}{Japanese journal of applied physics} \bibinfo{volume}{46}, \bibinfo{pages}{1482}.
\bibitem[{Gu et~al.(2018)Gu, He, Chen and Zhang}]{ref22}
\bibinfo{author}{Gu, Y.}, \bibinfo{author}{He, X.}, \bibinfo{author}{Chen, W.}, \bibinfo{author}{Zhang, C.}, \bibinfo{year}{2018}.
\newblock \bibinfo{title}{Analysis of three-dimensional anisotropic heat conduction problems on thin domains using an advanced boundary element method}.
\newblock \bibinfo{journal}{Computers \& Mathematics with Applications} \bibinfo{volume}{75}, \bibinfo{pages}{33--44}.
\bibitem[{Farrell et~al.(2005)Farrell, O’Riordan and Shishkin}]{ref23}
\bibinfo{author}{Farrell, P.}, \bibinfo{author}{O’Riordan, E.}, \bibinfo{author}{Shishkin, G.}, \bibinfo{year}{2005}.
\newblock \bibinfo{title}{A class of singularly perturbed semilinear differential equations with interior layers}.
\newblock \bibinfo{journal}{Mathematics of Computation} \bibinfo{volume}{74}, \bibinfo{pages}{1759--1776}.
\bibitem[{Zhang et~al.(2022)Zhang, Guo, Chen, Xu and Liu}]{ref24}
\bibinfo{author}{Zhang, Q.}, \bibinfo{author}{Guo, X.}, \bibinfo{author}{Chen, X.}, \bibinfo{author}{Xu, C.}, \bibinfo{author}{Liu, J.}, \bibinfo{year}{2022}.
\newblock \bibinfo{title}{Pinn-ffht: A physics-informed neural network for solving fluid flow and heat transfer problems without simulation data}.
\newblock \bibinfo{journal}{International Journal of Modern Physics C} \bibinfo{volume}{33}, \bibinfo{pages}{2250166}.
\bibitem[{LeCun et~al.(1998)LeCun, Bottou, Bengio and Haffner}]{ref25}
\bibinfo{author}{LeCun, Y.}, \bibinfo{author}{Bottou, L.}, \bibinfo{author}{Bengio, Y.}, \bibinfo{author}{Haffner, P.}, \bibinfo{year}{1998}.
\newblock \bibinfo{title}{Gradient-based learning applied to document recognition}.
\newblock \bibinfo{journal}{Proceedings of the IEEE} \bibinfo{volume}{86}, \bibinfo{pages}{2278--2324}.
\bibitem[{Raissi et~al.(2019)Raissi, Perdikaris and Karniadakis}]{ref26}
\bibinfo{author}{Raissi, M.}, \bibinfo{author}{Perdikaris, P.}, \bibinfo{author}{Karniadakis, G.E.}, \bibinfo{year}{2019}.
\newblock \bibinfo{title}{Physics-informed neural networks: A deep learning framework for solving forward and inverse problems involving nonlinear partial differential equations}.
\newblock \bibinfo{journal}{Journal of Computational physics} \bibinfo{volume}{378}, \bibinfo{pages}{686--707}.
\bibitem[{Kinga et~al.(2015)Kinga, Adam et~al.}]{ref27}
\bibinfo{author}{Kinga, D.}, \bibinfo{author}{Adam, J.B.}, et~al., \bibinfo{year}{2015}.
\newblock \bibinfo{title}{A method for stochastic optimization}, in: \bibinfo{booktitle}{International conference on learning representations (ICLR)}, \bibinfo{organization}{San Diego, California;}.
\bibitem[{Woltmann et~al.(2023)Woltmann, Thiessat, Hartmann, Habich and Lehner}]{ref28}
\bibinfo{author}{Woltmann, L.}, \bibinfo{author}{Thiessat, J.}, \bibinfo{author}{Hartmann, C.}, \bibinfo{author}{Habich, D.}, \bibinfo{author}{Lehner, W.}, \bibinfo{year}{2023}.
\newblock \bibinfo{title}{Fastgres: Making learned query optimizer hinting effective}.
\newblock \bibinfo{journal}{Proceedings of the VLDB Endowment} \bibinfo{volume}{16}, \bibinfo{pages}{3310--3322}.
\bibitem[{Jen and Anagonye(2001)}]{ref1}
\bibinfo{author}{Jen, T.C.}, \bibinfo{author}{Anagonye, A.U.}, \bibinfo{year}{2001}.
\newblock \bibinfo{title}{An improved transient model of tool temperatures in metal cutting}.
\newblock \bibinfo{journal}{J. Manuf. Sci. Eng.} \bibinfo{volume}{123}, \bibinfo{pages}{30--37}.
\bibitem[{Zhang(2018)}]{refAdam}
\bibinfo{author}{Zhang, Z.}, \bibinfo{year}{2018}.
\newblock \bibinfo{title}{Improved adam optimizer for deep neural networks}, in: \bibinfo{booktitle}{2018 IEEE/ACM 26th international symposium on quality of service (IWQoS)}, \bibinfo{organization}{IEEE}. pp. \bibinfo{pages}{1--2}.
\bibitem[{Lau and Lim(2018)}]{Tanh}
\bibinfo{author}{Lau, M.M.}, \bibinfo{author}{Lim, K.H.}, \bibinfo{year}{2018}.
\newblock \bibinfo{title}{Review of adaptive activation function in deep neural network}, in: \bibinfo{booktitle}{2018 IEEE-EMBS Conference on Biomedical Engineering and Sciences (IECBES)}, \bibinfo{organization}{IEEE}. pp. \bibinfo{pages}{686--690}.
\bibitem[{Paszke et~al.(2019)Paszke, Gross, Massa, Lerer, Bradbury, Chanan, Killeen, Lin, Gimelshein, Antiga et~al.}]{pytorch}
\bibinfo{author}{Paszke, A.}, \bibinfo{author}{Gross, S.}, \bibinfo{author}{Massa, F.}, \bibinfo{author}{Lerer, A.}, \bibinfo{author}{Bradbury, J.}, \bibinfo{author}{Chanan, G.}, \bibinfo{author}{Killeen, T.}, \bibinfo{author}{Lin, Z.}, \bibinfo{author}{Gimelshein, N.}, \bibinfo{author}{Antiga, L.}, et~al., \bibinfo{year}{2019}.
\newblock \bibinfo{title}{Pytorch: An imperative style, high-performance deep learning library}.
\newblock \bibinfo{journal}{Advances in neural information processing systems} \bibinfo{volume}{32}.

\end{thebibliography}



\end{document}